\documentclass[journal]{IEEEtran}

\usepackage{amsmath,bm,amssymb,amsfonts,amsthm}
\usepackage{algorithm,algorithmic}
\usepackage{graphicx}
\usepackage{xcolor}
\usepackage{relsize}
\usepackage{textcomp}
\usepackage{listings}

\usepackage{changepage}
\usepackage{hhline}
\usepackage{multirow}
\usepackage{nomencl}
\usepackage{mathtools}
\usepackage{commath}
\usepackage{booktabs}
\usepackage{subfiles}
\usepackage{tabularx}
\usepackage{lscape}
\usepackage{rotating}

\usepackage[normalem]{ulem}
\usepackage[utf8]{inputenc}
\usepackage[hyphens]{url}
\usepackage{lineno}
    \modulolinenumbers[5]
\usepackage[caption=false,font=footnotesize]{subfig}
\usepackage{float}
\usepackage{enumitem}
    \setlist{nolistsep}
\usepackage{scalerel}
\usepackage{tikz}
\usetikzlibrary{svg.path}
\definecolor{orcidlogocol}{HTML}{A6CE39}
\tikzset{
  orcidlogo/.pic={
    \fill[orcidlogocol] svg{M256,128c0,70.7-57.3,128-128,128C57.3,256,0,198.7,0,128C0,57.3,57.3,0,128,0C198.7,0,256,57.3,256,128z};
    \fill[white] svg{M86.3,186.2H70.9V79.1h15.4v48.4V186.2z}
                 svg{M108.9,79.1h41.6c39.6,0,57,28.3,57,53.6c0,27.5-21.5,53.6-56.8,53.6h-41.8V79.1z M124.3,172.4h24.5c34.9,0,42.9-26.5,42.9-39.7c0-21.5-13.7-39.7-43.7-39.7h-23.7V172.4z}
                 svg{M88.7,56.8c0,5.5-4.5,10.1-10.1,10.1c-5.6,0-10.1-4.6-10.1-10.1c0-5.6,4.5-10.1,10.1-10.1C84.2,46.7,88.7,51.3,88.7,56.8z};https://www.overleaf.com/project/5fbc1ff48658ea3b2cf56b6c
  }
}
\newcommand\orcidicon[1]{\href{https://orcid.org/#1}{\mbox{\scalerel*{
\begin{tikzpicture}[yscale=-1,transform shape]
\pic{orcidlogo};
\end{tikzpicture}
}{|}}}}
\usepackage[bookmarksopen,bookmarksdepth=3]{hyperref}

\usepackage[noadjust]{cite}

\begin{document}

\title{\huge Dynamic State Estimation for Radial Microgrid Protection}

\author{
    Arthur~K.~Barnes $^{1}$\orcidicon{0000-0001-9718-3197} \IEEEmembership{Member, IEEE}
    and~Adam~Mate $^{1}$\orcidicon{0000-0002-5628-6509} \IEEEmembership{Member, IEEE}

\thanks{Manuscript submitted: December~14,~2020. 
%Revised: XXXXX~XX,~2020. Accepted: XXXXX~XX,~2020. Date of publication: XXXXX~XX,~2020. Date of current version: XXXXX~XX,~2020.
}

\thanks{$^{1}$ The authors are with the Advanced Network Science Initiative at Los Alamos National Laboratory, Los Alamos, NM 87544 USA. Email:\{abarnes, amate\}@lanl.gov.}

\thanks{Color versions of one or more of the figures in this paper are available online at https://ieeexplore.ieee.org.}

\thanks{LANL ANSI LA-UR-20-30126.}
%\thanks{Digital Object Identifier: XX.XXXX/XXXXXX}

}

\markboth{IEEE/IAS 57th Industrial \& Commercial Power Systems Technical Conference, April~2021}{}

\maketitle

\begin{abstract}
Microgrids are localized electrical grids with control capability that are able to disconnect from the traditional grid to operate autonomously. They strengthen grid resilience, help mitigate grid disturbances, and support a flexible grid by enabling the integration of distributed energy resources.
Given the likely presence of critical loads, the proper protection of microgrids is of vital importance; however, this is complicated in the case of inverter-interfaced microgrids where low fault currents preclude the use of conventional time-overcurrent protection.
This paper introduces and investigates the application of dynamic state estimation, a generalization of differential protection, for the protection of radial portions of microgrids (or distribution networks); both phasor-based and dynamic approaches are investigated for protection. It is demonstrated through experiments on three case-study systems that dynamic state estimation is capable of correctly identifying model parameters for both normal and faulted operation.
\end{abstract}


\section{Introduction} \label{sec:intro}
\indent

Dynamic state estimation (abbr.~DSE) is a generalization of differential protection, which offers a reduced likelihood of misoperation, particularly in the case of assets with nonlinear characteristics (e.g., transformers that are being energized) \cite{meliopoulos_dynamic_2017}. It is also useful in cases where distance protection performs poorly (e.g., transmission lines with series compensation \cite{liu_dynamic_2017} or mutually coupled transmission lines \cite{liu_dynamic_2016}).
DSE has previously been introduced to microgrid branch protection \cite{Liu2015b, Vasios2018, Choi2017}.

This paper investigates the application of DSE for the protection of radial portions of a microgrid (or a distribution network).
This can be a challenge in electrical grids with distributed generation on account of lack of fault current from inverter-interfaced generation \cite{tumilty_approaches_2006}, varying fault current between grid-connected and islanded modes \cite{tumilty_approaches_2006}, and the potential for normally-meshed operation \cite{dewadasa_line_2008} and unbalanced operation due to single-phase loads \cite{dewadasa_line_2008}.
Admittance relaying has been investigated as a solution for the protection of microgrids \cite{barnes21-admitrelay}, however, it has been observed to present issues with grounded-wye connected loads \cite{dewadasa_distance_2008}; consequently, additional relaying is necessary to prevent misoperation \cite{dewadasa_line_2008}.

This paper treats the radial portions of a microgrid as load busses. It is assumed that these portions contain no loops or downstream generation; they are modeled as constant-impedance networks with unknown impedances but known connectivity.
To ensure that the number of measured variables is greater than approximately 1.6 times the number of free parameters (where 1.6 is a commonly selected number for redundancy to ensure sufficient measurements for system identification \cite{Kothari1989, Monticelli2000}), most models presented here make the assumption that the loads are balanced.

Every load and fault configuration requires a separate model. For a given load configuration, a model for each fault configuration is fit to measured values; the model with the lowest error, in terms of fitting the observed variables, is assumed to be the correct one.
On a grounded-wye-connected load, the following models are necessary to distinguish between normal operation, line-ground faults and line-line faults:
\begin{enumerate}
\item Normal operation: each branch of the load has the same impedance, which is modeled as a series resistive-inductive (abbr.~RL) network.
\item Phase A-ground fault: the faulted branch A is modeled as a resistance, while the unfaulted branches B and C are modeled as series RL networks with equal parameters.
\item Phase B-ground fault: the faulted branch B is modeled as a resistance, while the unfaulted branches C and A are modeled as series RL networks with equal parameters.
\item Phase C-ground fault: the faulted branch C is modeled as a resistance, while the unfaulted branches A and B are modeled as series RL networks with equal parameters.
\item Phase A-B fault: the fault impedance is modeled as a resistance across the load terminals A and B, while each branch of the load is modeled as a series RL network.
\item Phase B-C fault: the fault impedance is modeled as a resistance across the load terminals B and C, while each branch of the load is modeled as a series RL network.
\item Phase C-A fault: the fault impedance is modeled as a resistance across the load terminals C and A, while each branch of the load is modeled as a series RL network.
\end{enumerate}

On a delta-connected system, the following models are necessary to distinguish between normal operation, line-ground and line-line faults:
\begin{enumerate}
\item Normal operation: each branch of the load has the same impedance, which is modeled as series RL network.
\item Phase A-ground fault: the fault impedance is modeled as a resistance between load terminal A and ground, while the load branches are modeled as series RL networks.
\item Phase B-ground fault: the fault impedance is modeled as a resistance between load terminal B and ground, while the load branches are modeled as series RL networks.
\item Phase C-ground fault: the fault impedance is modeled as a resistance between load terminal C and ground, while the load branches are modeled as series RL networks.
\item Phase A-B fault: the fault impedance is modeled as a resistance across the load terminals A and B, while the branches across load terminals B-C and C-A are modeled as series RL networks.
\item Phase B-C fault: the fault impedance is modeled as a resistance across the load terminals B and C, while the branches across load terminals C-A and A-B are modeled as series RL networks.
\item Phase C-A fault: the fault impedance is modeled as a resistance across the load terminals C and A, while the branches across load terminals A-B and B-C are modeled as series RL networks.
\end{enumerate}
\vspace{0.05in}

Both phasor-based and dynamic approaches are investigated for radial microgrid protection.
Section~\ref{sec:phasor} describes the implementation of phasor-based state estimation: it is conceptually similar to DSE but more straightforward to derive and implement as it only requires a single time period.
Section~\ref{sec:dynamic} describes the implementation of DSE.
Section~\ref{sec:experiments} describes how two different transient models of loads are developed as test cases and run to test both phasor and dynamic state estimation; next, Section~\ref{sec:results} presents the performance of state estimation on the test cases.
Finally, Section~\ref{sec:conclusions} summarizes the conclusions of this paper.

\section{Phasor Implementation} \label{sec:phasor}
\indent

The phasor implementation of state estimation-based protection is simpler: only a single time period is used, which limits the number of measurements and therefore the number of parameters that can be estimated.
In this section, it is applied to single-phase, grounded-wye and delta-connected load configurations.

\subsection{Single-Phase Impedance}

The output of the system (illustrated in Fig.~\ref{fig:rl-phasor}):
\begin{equation*}
\mathbf{y} = \begin{bmatrix} V \\ I \end{bmatrix}
\end{equation*}
\noindent where $V$ and $I$ are phasor quantities.
\vspace{0.1in}

The state of the system:
\begin{equation*}
\mathbf{x} = \begin{bmatrix} Z \\ I_z \end{bmatrix}
\end{equation*}

The output-state mapping for the system is the following vector-valued function:
\begin{equation*}
\mathbf{y} = \mathbf{h}(\mathbf{x})
\end{equation*}
\noindent where
\begin{align*}
h_1(\mathbf{x}) & = V_z = Z I_z &
h_2(\mathbf{x}) & = I_z
\end{align*}

The Jacobian of $\mathbf{h}(\mathbf{x})$ is determined as follows:
\begin{align*}
\frac{\partial V_z}{\partial Z} & = \frac{\partial}{\partial Z} Z I_z = I_z &
\frac{\partial V_z}{\partial I_z} & = \frac{\partial}{\partial I_z} Z I_z = Z \\
\frac{\partial I_z}{\partial Z} & = \frac{\partial}{\partial Z} I_z = 0 &
\frac{\partial I_z}{\partial I_z} & = \frac{\partial}{\partial I_z} I_z = 1 
\end{align*}

\noindent The mapping between variables and the state vector:
\begin{align*}
Z & = x_1 &
I_z & = x_2
\end{align*}

\noindent Given the variable and state mapping, the Jacobian can be built as follows: $H(n,m)= 0$, unless specified below.
\begin{equation*}
H = 
\begin{bmatrix}[1.5]
\frac{\partial V_z}{\partial Z} & \frac{\partial V_z}{\partial I_z} \\
\frac{\partial I_z}{\partial Z} & \frac{\partial I_z}{\partial I_z}
\end{bmatrix}
\end{equation*}

\noindent Given the Jacobian, the state of the system can be solved for iteratively:
\begin{align*}
{\boldsymbol\epsilon}_i & = \mathbf{y} - \mathbf{h}(\mathbf{x}_i) \hspace{0.5in} &
J_i & = ||{\boldsymbol\epsilon}_i||^2 \\
\mathbf{x}_{i+1} & = \mathbf{x}_i + (H'_i H_i)^{-1}H'_i {\boldsymbol\epsilon}_i
\end{align*}

\subsection{Grounded-Wye with Line-Ground Fault}

The output of the system (illustrated in Fig.~\ref{fig:gwye-phasor}):
\begin{equation*}
\mathbf{y} = \begin{bmatrix} I_a & I_b & I_c & V_a & V_b & V_c \end{bmatrix}^T
\end{equation*}

The easiest way to model this is as an unbalanced load where the fault impedance is not treated specially.
The state of the system therefore:
\begin{equation*}
\mathbf{x} = \begin{bmatrix} Y_a & Y_b & Y_c & V_{za} & V_{zb} & V_{zc} \end{bmatrix}^T
\end{equation*}

The output function $\mathbf{h}(\mathbf{x})$ can be written as:
\begin{align*}
\mathbf{h}_1(\mathbf{x}) & = I_a = y_aV_a &
\mathbf{h}_2(\mathbf{x}) & = I_b = y_bV_b \\
\mathbf{h}_3(\mathbf{x}) & = I_c = y_cV_c &
\mathbf{h}_4(\mathbf{x}) & = V_a = V_{za} \\
\mathbf{h}_5(\mathbf{x}) & = V_b = V_{zb} &
\mathbf{h}_6(\mathbf{x}) & = V_c = V_{zc}
\end{align*}

\subsection{Grounded-Wye with Line-Line Fault}

The output of the system (illustrated in Fig.~\ref{fig:gwye-llf-phasor}):
\begin{equation*}
\mathbf{y} = \begin{bmatrix} I_a & I_b & I_c & V_a & V_b & V_c \end{bmatrix}^T
\end{equation*}

The state of the system:
\begin{equation*}
\mathbf{x} = \begin{bmatrix} Y_l & Y_f & V_{za} & V_{zb} & V_{zc} \end{bmatrix}^T
\end{equation*}

The output function $\mathbf{h}(\mathbf{x})$ can be written as:
\begin{align*}
h_1(\mathbf{x}) & = I_a = (y_l + y_f)V_{za} - y_fV_{zb} \\
h_2(\mathbf{x}) & = I_b = -y_fV_{za} + y_lV_{zb} \\
h_3(\mathbf{x}) & = I_c = y_lV_{zc} \\
h_4(\mathbf{x}) & = V_a = V_{za} \\
h_5(\mathbf{x}) & = V_b = V_{zb} \\
h_6(\mathbf{x}) & = V_c = V_{zc}
\end{align*}

\subsection{Delta-Connected Load with Line-Line Fault}

The output of the system (illustrated in Fig.~\ref{fig:delta-llf-phasor}):
\begin{equation*}
\mathbf{y} = \begin{bmatrix} I_a & I_b & I_c & V_a & V_b & V_c \end{bmatrix}^T
\end{equation*}

The state of the system:
\begin{equation*}
\mathbf{x} = \begin{bmatrix} Y_f & Y_{ll} & V_{za} & V_{zb} & V_{zc} \end{bmatrix}^T
\end{equation*}

The output function $\mathbf{h}(\mathbf{x})$ can be written as:
\begin{align*}
h_1(\mathbf{x}) & = I_a = y_{aa}V_{za} - y_{ab}V_{zb} - y_{ca}V_{zc} \\
& = (y_{ab} + y_{ca})V_{za} - y_{ab}V_{zb} - y_{ca}V_{zc}  \\
& = (y_f + y_{ll})V_{za} - y_fV_{sb} - y_{ll}V_{zc} \\
h_2(\mathbf{x}) & = I_b = -y_{ab}V_{za} + y_{bb}V_{zb} - y_{bc}V_{zc} \\
& = -y_{ab}V_{za} + (y_{ab} + y_{bc})V_{zb} - y_{bc}V_{zc} \\
& = -y_fV_{za} + (y_f + y_{ll})V_{zb} - y_{ll}V_{zc} \\
h_3(\mathbf{x}) & = I_c = -y_{ca}V_{za} - y_{bc}V_{zb} + y_{cc}V_{zc} \\
& =  -y_{ca}V_{za} - y_{bc}V_{zb} + (y_{ac} + y_{bc})V_{zc} \\
& = -y_{ll}V_{za} - y_{ll}V_{zb} + 2y_{ll}V_{zc} \\
h_4(\mathbf{x}) & = V_a = V_{za} \\
h_5(\mathbf{x}) & = V_b = V_{zb} \\
h_6(\mathbf{x}) & = V_c = V_{zc}
\end{align*}

\subsection{Delta-Connected Load with a Line-Ground Fault}

\begin{figure*}[!htbp]
\centering

\subfloat[Single-phase load]{\includegraphics[width=1.25in]{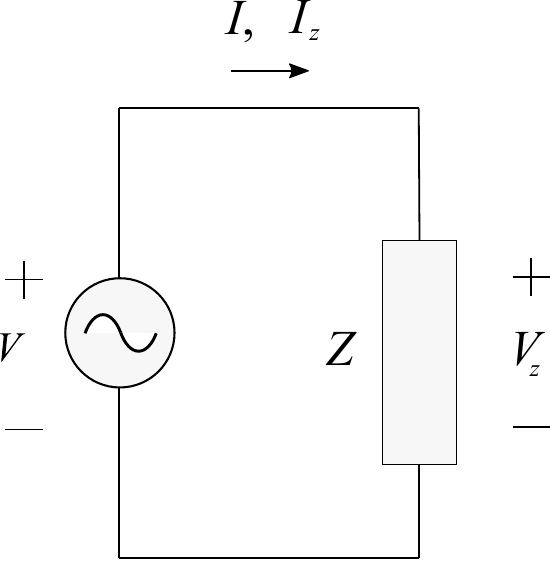}
\label{fig:rl-phasor}}
\hfil
\subfloat[Grounded-wye load with line-ground fault fault]{\includegraphics[width=2.55in]{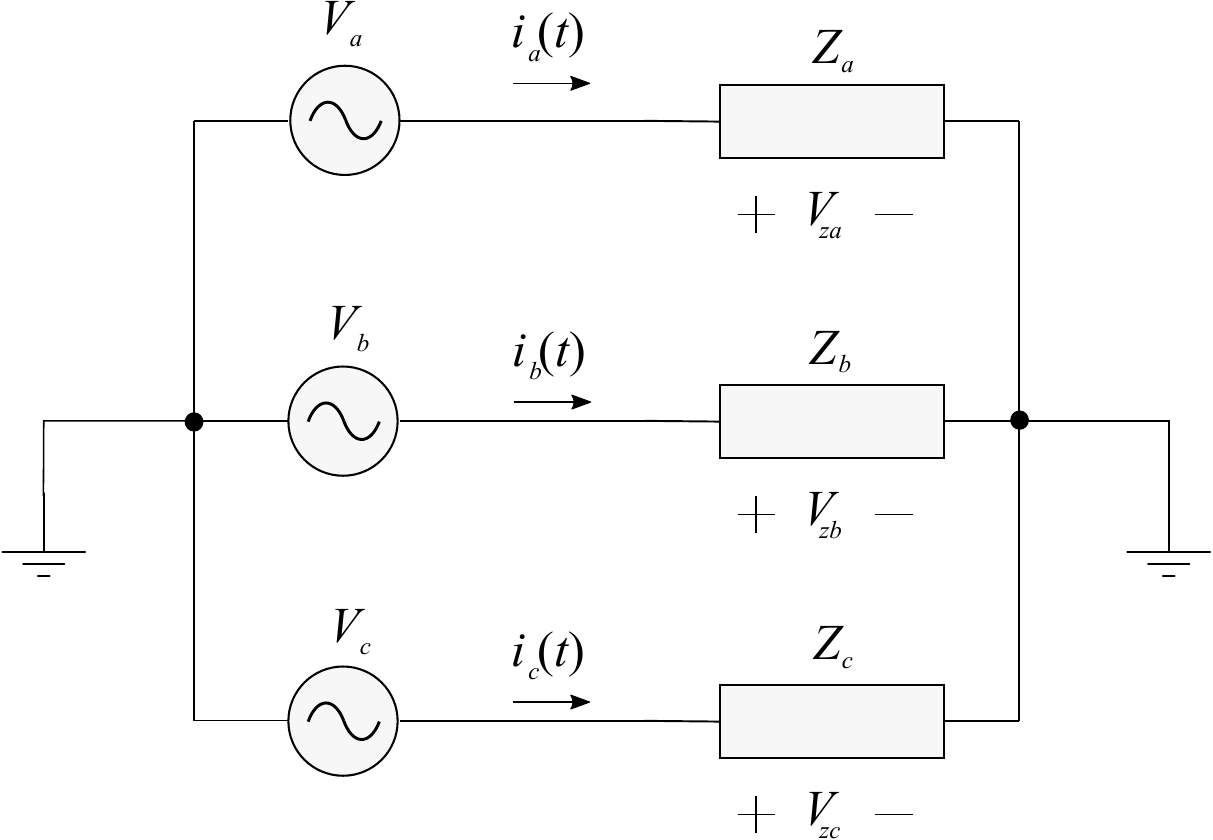}
\label{fig:gwye-phasor}}
\hfil
\subfloat[Grounded-wye load with line-line fault]{\includegraphics[width=2.55in]{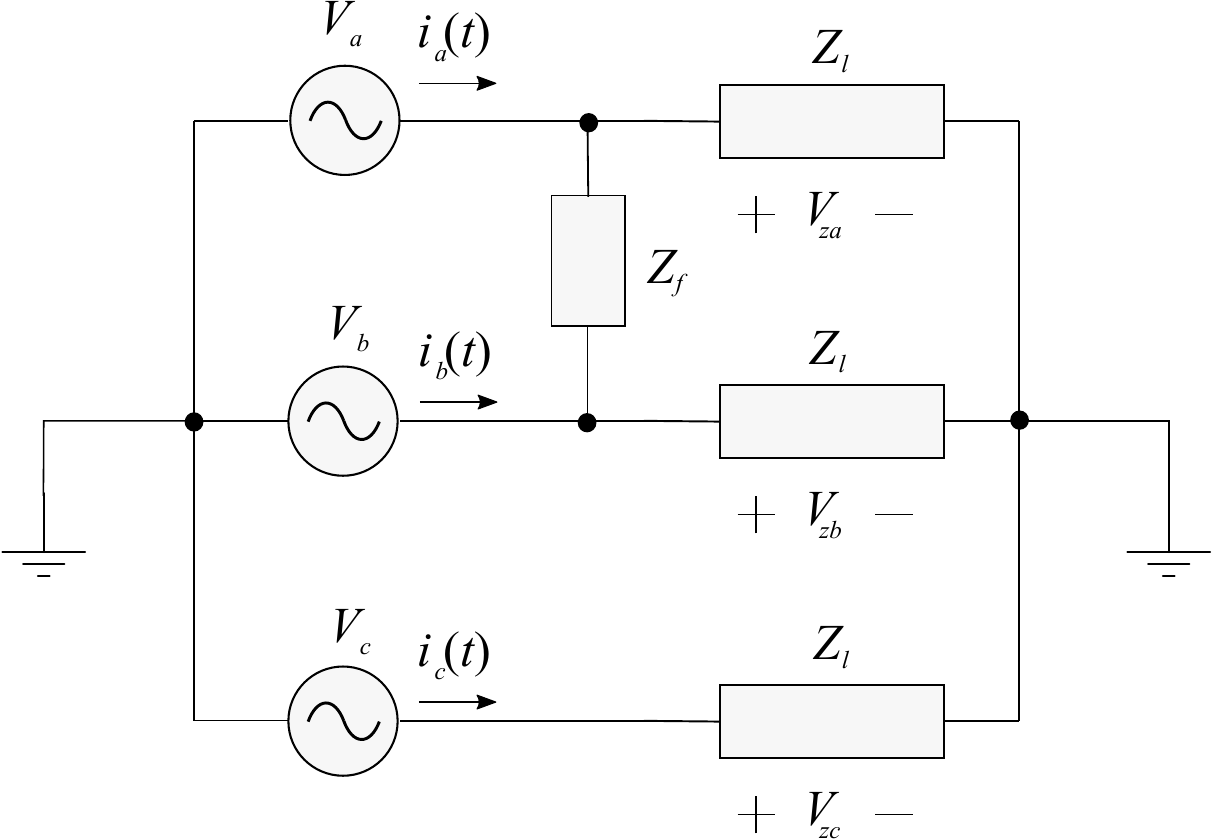}
\label{fig:gwye-llf-phasor}}

\subfloat[Delta-connected load with a line-line fault]{\includegraphics[width=2.25in]{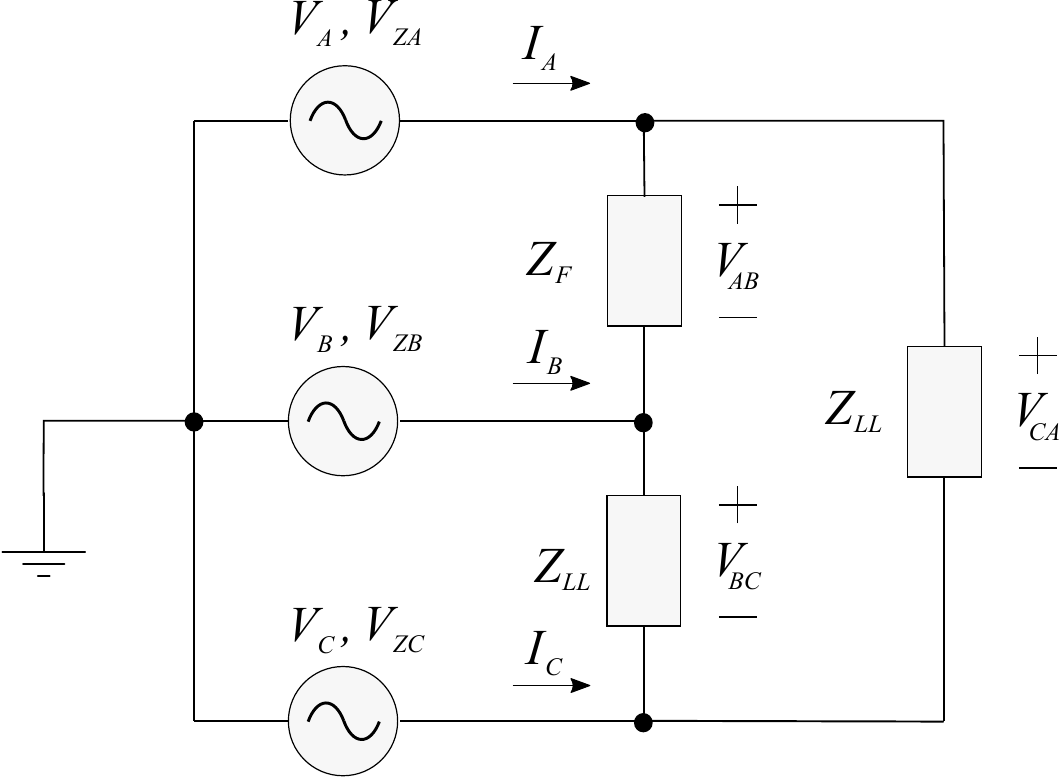}
\label{fig:delta-llf-phasor}}
\hfil
\subfloat[Delta-connected load with a line-ground fault]{\includegraphics[width=2.85in]{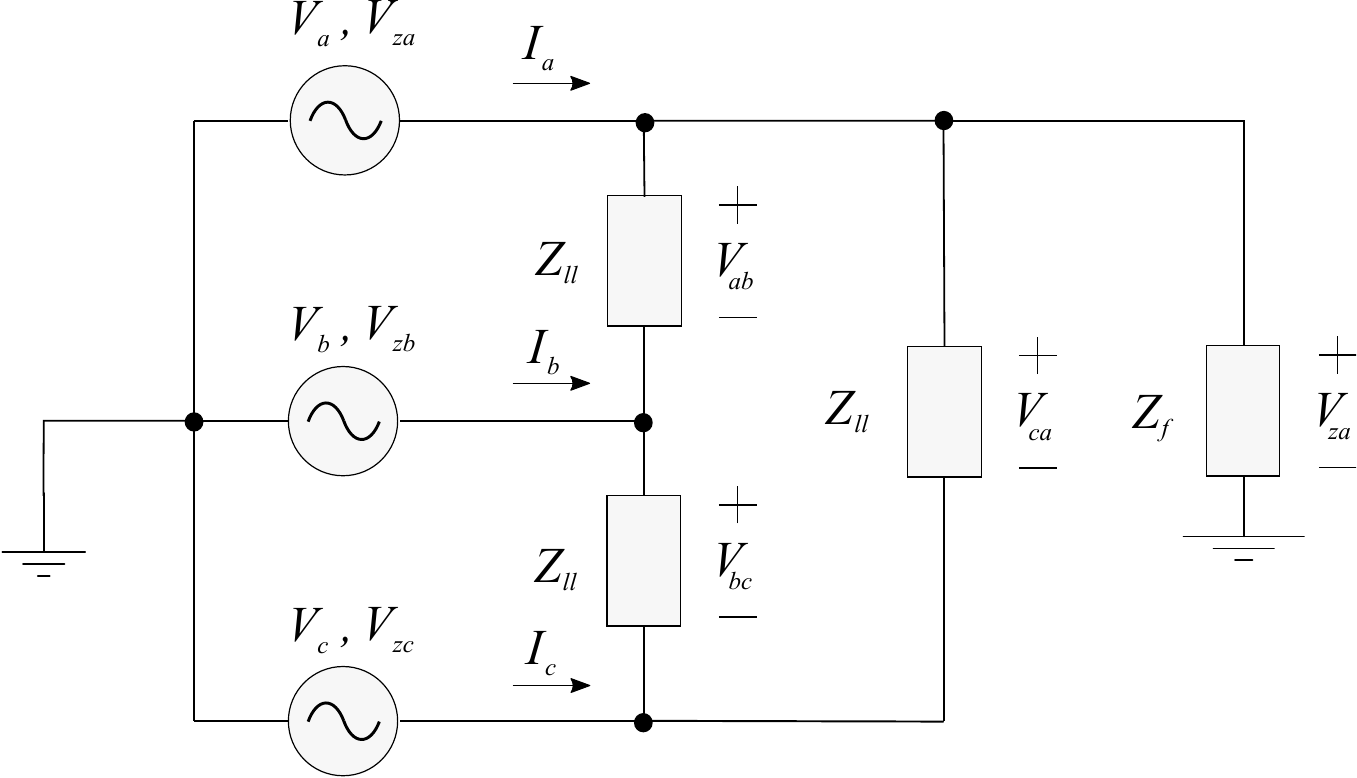}
\label{fig:delta-lgf-phasor}}

\vspace{0.1in}
\caption{Phasor-based implementation of state estimation-based protection, applied to single-phase, grounded-wye and delta-connected load configurations.}
\label{fig:phasor-models}
\end{figure*}

The output of the system (illustrated in Fig.~\ref{fig:delta-lgf-phasor}):
\begin{equation*}
\mathbf{y} = \begin{bmatrix} I_a & I_b & I_c & V_a & V_b & V_c \end{bmatrix}^T
\end{equation*}

The state of the system:
\begin{equation*}
\mathbf{x} = \begin{bmatrix} Y_{ll} & Y_f & V_{za} & V_{zb} & V_{zc} \end{bmatrix}^T
\end{equation*}

The output function $\mathbf{h}(\mathbf{x})$ can be written as:
\begin{align*}
h_1(\mathbf{x}) & = I_a = y_{aa}V_{za} - y_{ab}V_{zb} - y_{ca}V_{zc} \\
& = (y_{ab} + y_{ca} + y_{ag})V_{za} - y_{ab}V_{zb} - y_{ca}V_{zc} \\
& = (y_f + 2y_{ll})V_{za} - y_{ll}V_{sb} - y_{ll}V_{zc} \\
h_2(\mathbf{x}) & = I_b = -y_{ab}V_{za} + y_{bb}V_{zb} - y_{bc}V_{zc} \\
& = -y_{ab}V_{za} + (y_{ab} + y_{bc})V_{zb} - y_{bc}V_{zc} \\
& = -y_{ll}V_{za} + 2y_{ll}V_{zb} - y_{ll}V_{zc} \\
h_3(\mathbf{x}) & = I_c = -y_{ca}V_{za} - y_{bc}V_{zb} + y_{cc}V_{zc} \\
& =  -y_{ca}V_{za} - y_{bc}V_{zb} + (y_{ac} + y_{bc})V_{zc} \\
& = -y_{ll}V_{za} - y_{ll}V_{zb} + 2y_{ll}V_{zc} \\
h_4(\mathbf{x}) & = V_a = V_{za} \\
h_5(\mathbf{x}) & = V_b = V_{zb} \\
h_6(\mathbf{x}) & = V_c = V_{zc}
\end{align*}

\section{Dynamic Implementation} \label{sec:dynamic}
\indent

While the phasor implementation uses a single time period for state estimation, with the dynamic implementation several periods are used; in this paper, 12 cycles are sampled at a 2~[kHz] sample rate.
As in Section~\ref{sec:phasor}, here the DSE-based protection is applied to single-phase, grounded-wye and delta-connected load configurations.

\subsection{Single-Phase Series RL Load}

The output of the system (illustrated in Fig.~\ref{fig:rl-dynamic}):
\begin{equation*}
y(t) = \begin{bmatrix} v(t) \\ i(t) z(t) \end{bmatrix}
\end{equation*}

\noindent For the purposes of state estimation, this is sampled at points $n \in \{1, ..., N\}$ giving the following vector-value equation:
\begin{equation*}
\mathbf{y} = \begin{bmatrix} \mathbf{v} \\ \mathbf{i} \end{bmatrix}
\end{equation*}
\noindent where
\begin{align*}
\mathbf{v} & = \begin{bmatrix}v(1) & v(2) & \cdots & v(N)\end{bmatrix}^T \\
\mathbf{i} & = \begin{bmatrix}i(1) & i(2) & \cdots & i(N)\end{bmatrix}^T \\
\mathbf{z} & = \begin{bmatrix}z(1) & z(2) & \cdots & z(N)\end{bmatrix}^T
\end{align*}

The state of the system:
\begin{equation*}
\mathbf{x} = \begin{bmatrix} R & L & \mathbf{v}_r & \mathbf{v}_l \end{bmatrix}^T
\end{equation*}

The output-state mapping for the system is the following vector-valued function:
\begin{equation*}
\mathbf{y} = \mathbf{h}(\mathbf{x})
\end{equation*}
\noindent where
\begin{gather*}
h_n(\mathbf{x}) = v_r(n) + v_l(n) \quad \forall n \in \{1,2,\ldots,N\} \\
h_{N+n}(\mathbf{x}) = Gv_r(n) \quad \forall n \in \{1,2,\ldots,N\} \\
h_{2N+n}(\mathbf{x}) = Gv_r(n) - Gv_r(n-2) + \frac{2\Lambda\Delta t}{6}(v_l(n) + \\ + 4v_l(n-1) + v_l(n-2)), \quad \forall n \in \{3,4,\ldots,N\}
\end{gather*}
\vspace{0.05in}

\noindent In the above, $v_R(n) = R i_L(n)$ follows from discretizing $v_R(t) = R i_L(t)$, and $$v_l(n) = \frac{2\Lambda\Delta t}{6}(v_l(n) + 4v_l(n-1) + v_l(n-2))$$ follows from discretizing
$$i_l(t) = \frac{1}{L}\int_{t-\Delta t}^{t} v_l(\tau) d\tau$$ via Simpson's 1/3 rule \cite{Chapra2009}.
\vspace{0.1in}

Given the variable and state vector mapping, the Jacobian can be built as follows: $H(n,m)= 0$, unless specified otherwise below.
\begin{gather*}
H(n,2+n) = \frac{\partial v(n)}{\partial v_r(n)} = 1 \quad \forall n \in \{1, 2, \ldots, N\} \\
H(n,2+N+n) = \frac{\partial v(n)}{\partial v_l(n)} = 1 \quad \forall n \in \{1, 2, \ldots, N\} \\
H(N+n,1) = \frac{\partial i(n)}{\partial G} = v_r(n) \quad \forall n \in \{1, 2, \ldots, N\} \\
H(N+n,2+n) = \frac{\partial i(n)}{\partial v_r(n)} = G \quad \forall n \in \{1, 2, \ldots, N\} \\
H(2N+n-2,1) = \frac{\partial z(n-2)}{\partial R} = v_r(n) - v_r(n-2) \\ \quad \forall n \in \{3, 4, \ldots, N\} \\
\end{gather*}
\begin{gather*}
H(2N+n-2,n) = \frac{\partial z(n-2)}{\partial v_r(n)} = G \quad \forall n \in \{3, 4, \ldots, N\} \\
H(2N+n-2,2) = \frac{\partial z(n-2)}{\partial \Lambda} = \frac{2\Delta t}{6}(v_l(n) + \\ + 4v_l(n-1) + v_l(n-2)) \quad \forall n \in \{3, 4, \ldots, N\} \\
H(2N+n-2,2+N+n) = \frac{\partial z(n-2)}{\partial v_l(n)} = \frac{\Delta t \Lambda}{3}  \\ \quad \forall n \in \{1, 2, \ldots, N\} \\
H(2N+n-2,1+N+n) = \frac{\partial z(n-2)}{\partial v_l(n-1)} = \frac{4\Delta t \Lambda}{3} \\ \quad \forall n \in \{1, 2, \ldots, N\} \\
H(2N+n-2,N+n) = \frac{\partial z(n-2)}{\partial v_l(n-2)} = \frac{\Delta t \Lambda}{3} \\ \quad \forall n \in \{1, 2, \ldots, N\}
\end{gather*}

\noindent Given the Jacobian, the state of the system can be solved for iteratively, by applying the same equations as in Section~II-A.

\subsection{Grounded-Wye Load without Fault}

The sampled output of the system (illustrated in Fig.~\ref{fig:gwye-no-fault-dynamic}):
\begin{equation*}
\mathbf{y} = \begin{bmatrix} \mathbf{v}_a & \mathbf{v}_b & \mathbf{v}_c & \mathbf{i}_a & \mathbf{i}_b & \mathbf{i}_c & \mathbf{z}_a & \mathbf{z}_b & \mathbf{z}_c \end{bmatrix}^T
\end{equation*}
\noindent where
\begin{align*}
\mathbf{v}_\phi & = \begin{bmatrix}v_\phi(1) & v_\phi(2) & \cdots & v_\phi(N)\end{bmatrix}^T \\
\mathbf{i}_\phi & = \begin{bmatrix}i_\phi(1) & i_\phi(2) & \cdots & i_\phi(N)\end{bmatrix}^T \\
\mathbf{z}_\phi & = \begin{bmatrix}z_\phi(1) & z_\phi(2) & \cdots & z_\phi(N-2)\end{bmatrix}^T for \hspace{0.05in} \phi \in \{a,b,c\}
\end{align*}

\noindent The state of the system:
\begin{equation*}
x(t) = \begin{bmatrix} G & \Lambda & \mathbf{v}_{ra} & \mathbf{v}_{rb} & \mathbf{v}_{rc} & \mathbf{v}_{la} & \mathbf{v}_{lb} & \mathbf{v}_{lc} \end{bmatrix}^T
\end{equation*}
\noindent where $G = R^{-1}$ is the conductance, $\Lambda = L^{-1}$ is the reciprocal of the inductance, $\mathbf{v}_{r\phi}$ is the voltage across the resistance on phase $\phi$ at each time period $1, \ldots, N$ and $\mathbf{v}_{l\phi}$ is the voltage across the inductance on phase $\phi$ at each time period $1, \ldots, N$.
\vspace{0.1in}

The output function $\mathbf{h}(\mathbf{x})$ can be written as:
\begin{gather*}
h_n(\mathbf{x}) = v_{ra}(n) + v_{la}(n) \quad \forall n \in \{1, 2, \ldots, N\} \\
h_{n+N}(\mathbf{x}) = v_{rb}(n) + v_{lb}(n) \quad \forall n \in \{1, 2, \ldots, N\} \\
h_{n+2N}(\mathbf{x}) = v_{rb}(n) + v_{lb}(n) \quad \forall n \in \{1, 2, \ldots, N\} \\
h_{n+3N}(\mathbf{x}) = Gv_{ra}(n) \quad \forall n \in \{1, 2, \ldots, N\} \\
h_{n+4N}(\mathbf{x}) = Gv_{rb}(n) \quad \forall n \in \{1, 2, \ldots, N\} \\
h_{n+5N}(\mathbf{x}) = Gv_{rc}(n)) \quad \forall n \in \{1, 2, \ldots, N\} \\    
h_{n+6N}(\mathbf{x}) = G(v_{ra}(n) - v_{ra}(n-2)) - \frac{2\Delta t\Lambda}{6}(v_{la}(n) + \\ + 4v_{la}(n-1) + v_{la}(n-2)) \quad \forall n \in \{1, 2, \ldots, N\} \\
h_{n+7N}(\mathbf{x}) = G(v_{rb}(n) - v_{rb}(n-2)) - \frac{2\Delta t\Lambda}{6}(v_{lb}(n) + \\ + 4v_{lc}(n-1) + v_{lb}(n-2)) \quad \forall n \in \{1, 2, \ldots, N\} \\
h_{n+8N}(\mathbf{x}) = G(v_{rc}(n) - v_{rc}(n-2)) - \frac{2\Delta t\Lambda}{6}(v_{lc}(n) + \\ + 4v_{lc}(n-1) + v_{lc}(n-2)) \quad \forall n \in \{1, 2, \ldots, N\}.   
\end{gather*}

\subsection{Grounded-Wye Load with Line-Ground Fault}

The sampled output of the system (illustrated in Fig.~\ref{fig:gwye-lg-fault-r-only-dynamic}):
\begin{equation*}
\mathbf{y} = \begin{bmatrix} \mathbf{v}_a & \mathbf{v}_b & \mathbf{v}_c & \mathbf{i}_a & \mathbf{i}_b & \mathbf{i}_c &\mathbf{z}_b & \mathbf{z}_c \end{bmatrix}^T
\end{equation*}
\noindent Note that there are no $z_a(n)$ output variables as the reactive impedance on phase A is large compared to the parallel fault conductance $G_f$.
\vspace{0.1in}

The state of the system:
\begin{equation*}
\label{eq:gwye-lg-fault-x}
x(t) = \begin{bmatrix} G & \Lambda & G_f & \mathbf{v}_{ra} & \mathbf{v}_{rb} & \mathbf{v}_{rc} & \mathbf{v}_{lb} & \mathbf{v}_{lc} \end{bmatrix}^T
\end{equation*}
\noindent where $G_f = R^{-1}$ is the conductance and the remaining states are the same as those in that of the grounded-wye no-fault state.
\vspace{0.1in}

The output function $\mathbf{h}(\mathbf{x})$ can be written as:
\begin{gather*}
h_n(\mathbf{x}) = v_{ra}(n) \quad \forall n \in \{1, 2, \ldots, N\} \\
h_{n+N}(\mathbf{x}) = v_{rb}(n) + v_{lb}(n) \quad \forall n \in \{1, 2, \ldots, N\} \\
h_{n+2N}(\mathbf{x}) = v_{rb}(n) + v_{lb}(n) \quad \forall n \in \{1, 2, \ldots, N\} \\
h_{n+3N}(\mathbf{x}) = Gv_{ra}(n) \quad \forall n \in \{1, 2, \ldots, N\} \\
h_{n+4N}(\mathbf{x}) = Gv_{rb}(n) \quad \forall n \in \{1, 2, \ldots, N\} \\
h_{n+5N}(\mathbf{x}) = Gv_{rc}(n)) \quad \forall n \in \{1, 2, \ldots, N\} \\    
h_{n+6N}(\mathbf{x}) = G(v_{rb}(n) - v_{rb}(n-2)) - \frac{2\Delta t\Lambda}{6}(v_{lb}(n) + \\ + 4v_{lc}(n-1) + v_{lb}(n-2)) \quad \forall n \in \{1, 2, \ldots, N\} \\
h_{n+7N}(\mathbf{x}) = G(v_{rc}(n) - v_{rc}(n-2)) - \frac{2\Delta t\Lambda}{6}(v_{lc}(n) + \\ + 4v_{lc}(n-1) + v_{lc}(n-2)) \quad \forall n \in \{1, 2, \ldots, N\}  
\end{gather*}

\subsection{Grounded-Wye Load with Line-Line Fault}

The sampled output of the system (illustrated in Fig.~\ref{fig:gwye-ll-fault-r-only-dynamic}):
\begin{equation*}
\mathbf{y} = \begin{bmatrix} \mathbf{v}_a & \mathbf{v}_b & \mathbf{v}_c & \mathbf{i}_a & \mathbf{i}_b & \mathbf{i}_c &\mathbf{z}_a & \mathbf{z}_b & \mathbf{z}_c \end{bmatrix}^T
\end{equation*}

The state of the system:
\begin{equation*}
x(t) = \begin{bmatrix} G & \Lambda & \mathbf{v}_{ra} & \mathbf{v}_{rb} & \mathbf{v}_{rc} & \mathbf{v}_{la} & \mathbf{v}_{lb} & \mathbf{v}_{lc} \end{bmatrix}^T
\end{equation*}

The output function $\mathbf{h}(\mathbf{x})$ can be written as:
\begin{gather*}
h_n(\mathbf{x}) = v_{ra}(n) + v_{la}(n) \quad \forall n \in \{1, 2, \ldots, N\} \\
h_{n+N}(\mathbf{x}) = v_{rb}(n) + v_{lb}(n) \quad \forall n \in \{1, 2, \ldots, N\} \\
h_{n+2N}(\mathbf{x}) = v_{rb}(n) + v_{lb}(n) \quad \forall n \in \{1, 2, \ldots, N\} \\
h_{n+3N}(\mathbf{x}) = Gv_{ra}(n) + G_f(v_{ra}(n) + v_{la}(n) - v_{rb}(n) - \\ - v_{lb}(n)) \quad \forall n \in \{1, 2, \ldots, N\} \\
h_{n+4N}(\mathbf{x}) = Gv_{rb}(n) - G_f(v_{ra}(n) + v_{la}(n) - v_{rb}(n) - \\ - v_{lb}(n)) \quad \forall n \in \{1, 2, \ldots, N\} \\
h_{n+5N}(\mathbf{x}) = Gv_{rc}(n)) \quad \forall n \in \{1, 2, \ldots, N\} \\    
h_{n+6N}(\mathbf{x}) = G(v_{ra}(n) - v_{ra}(n-2)) - \frac{2\Delta t\Lambda}{6}(v_{la}(n) + \\ + 4v_{la}(n-1) + v_{la}(n-2)) \quad \forall n \in \{1, 2, \ldots, N\} \\
\end{gather*}
\begin{gather*}
h_{n+7N}(\mathbf{x}) = G(v_{rb}(n) - v_{rb}(n-2)) - \frac{2\Delta t\Lambda}{6}(v_{lb}(n) + \\ + 4v_{lc}(n-1) + v_{lb}(n-2)) \quad \forall n \in \{1, 2, \ldots, N\} \\
h_{n+8N}(\mathbf{x}) = G(v_{rc}(n) - v_{rc}(n-2)) - \frac{2\Delta t\Lambda}{6}(v_{lc}(n) + \\ + 4v_{lc}(n-1) + v_{lc}(n-2)) \quad \forall n \in \{1, 2, \ldots, N\} \\    
\end{gather*}

\subsection{Delta Load without Fault}

The sampled output of the system (illustrated in Fig.~\ref{fig:delta-ll-fault-r-only-dynamic}):
\begin{equation*}
\mathbf{y} = \begin{bmatrix} \mathbf{v}_{ab} & \mathbf{v}_{bc} & \mathbf{v}_{ca} & \mathbf{i}_a & \mathbf{i}_b & \mathbf{i}_c, &\mathbf{z}_{ab} & \mathbf{z}_{bc} & \mathbf{z}_{ca} \end{bmatrix}^T
\end{equation*}

The state of the system:
\begin{equation*}
x(t) = \begin{bmatrix} G & \Lambda \mathbf{v}_{rab} & \mathbf{v}_{rbc} & \mathbf{v}_{rca} & \mathbf{v}_{lab} & \mathbf{v}_{lbc} & \mathbf{v}_{lca} \end{bmatrix}^T
\end{equation*}

The output function $\mathbf{h}(\mathbf{x})$ can be written as:
\begin{gather*}
h_n(\mathbf{x}) = v_{rab}(n) + v_{lab}(n) \quad \forall n \in \{1, 2, \ldots, N\} \\
h_{n+N}(\mathbf{x}) = v_{rbc}(n) + v_{lbc}(n) \quad \forall n \in \{1, 2, \ldots, N\} \\
h_{n+2N}(\mathbf{x}) = v_{rca}(n) + v_{lca}(n) \quad \forall n \in \{1, 2, \ldots, N\} \\
h_{n+3N}(\mathbf{x}) = G(v_{rab}(n) - v_{rca}(n)) \quad \forall n \in \{1, 2, \ldots, N\} \\
h_{n+4N}(\mathbf{x}) = G(v_{rbc}(n) - v_{rab}(n))\quad \forall n \in \{1, 2, \ldots, N\} \\
h_{n+5N}(\mathbf{x}) = G(v_{rca}(n) - v_{rbc}(n)) \quad \forall n \in \{1, 2, \ldots, N\} \\    
h_{n+6N}(\mathbf{x}) = G(v_{rab}(n) - v_{rab}(n-2)) - \frac{2\Delta t\Lambda}{6}(v_{lab}(n) + \\ + 4v_{lab}(n-1) + v_{lab}(n-2)) \quad \forall n \in \{1, 2, \ldots, N\} \\
h_{n+7N}(\mathbf{x}) = G(v_{rbc}(n) - v_{rbc}(n-2)) - \frac{2\Delta t\Lambda}{6}(v_{lbc}(n) + \\ + 4v_{lbc}(n-1) + v_{lbc}(n-2)) \quad \forall n \in \{1, 2, \ldots, N\} \\
h_{n+8N}(\mathbf{x}) = G(v_{rca}(n) - v_{rca}(n-2)) - \frac{2\Delta t\Lambda}{6}(v_{lca}(n) + \\ + 4v_{lca}(n-1) + v_{lca}(n-2)) \quad \forall n \in \{1, 2, \ldots, N\} \\    
\end{gather*}

\subsection{Delta Load with Line-Line Fault}

The sampled output of the system (illustrated in Fig.~\ref{fig:gwye-ll-fault-r-only-dynamic}):
\begin{equation*}
\mathbf{y} = \begin{bmatrix} \mathbf{v}_{ab} & \mathbf{v}_{bc} & \mathbf{v}_{ca} & \mathbf{i}_a & \mathbf{i}_b & \mathbf{i}_c, &\mathbf{z}_{bc} & \mathbf{z}_{ca} \end{bmatrix}^T
\end{equation*}
\noindent Note that there are no $z_{ab}(n)$ output variables as the reactive impedance on phase A is large compared to the parallel fault conductance $G_f$.
\vspace{0.1in}

The state of the system:
\begin{equation*}
x(t) = \begin{bmatrix} G & \Lambda & G_f & \mathbf{v}_{rab} & \mathbf{v}_{rbc} & \mathbf{v}_{rca} & \mathbf{v}_{lbc} & \mathbf{v}_{lca} \end{bmatrix}^T
\end{equation*}
\noindent where $G_f = R^{-1}$ is the conductance and the remaining states are the same as those in that of the delta no-fault state.
\vspace{0.1in}

The output function $\mathbf{h}(\mathbf{x})$ can be written as:
\begin{gather*}
h_n(\mathbf{x}) = v_{rab}(n) \quad \forall n \in \{1, 2, \ldots, N\} \\
h_{n+N}(\mathbf{x}) = v_{rbc}(n) + v_{lbc}(n) \quad \forall n \in \{1, 2, \ldots, N\} \\
h_{n+2N}(\mathbf{x}) = v_{rca}(n) + v_{lca}(n) \quad \forall n \in \{1, 2, \ldots, N\} \\
h_{n+3N}(\mathbf{x}) = Gv_{ra}(n) \quad \forall n \in \{1, 2, \ldots, N\} \\
h_{n+4N}(\mathbf{x}) = Gv_{rb}(n) \quad \forall n \in \{1, 2, \ldots, N\} \\
\end{gather*}
\begin{gather*}
h_{n+5N}(\mathbf{x}) = Gv_{rc}(n)) \quad \forall n \in \{1, 2, \ldots, N\} \\
h_{n+6N}(\mathbf{x}) = G(v_{rbc}(n) - v_{rbc}(n-2)) - \frac{2\Delta t\Lambda}{6}(v_{lbc}(n) + \\ + 4v_{lbc}(n-1) + v_{lbc}(n-2)) \quad \forall n \in \{1, 2, \ldots, N\} \\
h_{n+7N}(\mathbf{x}) = G(v_{rca}(n) - v_{rca}(n-2)) - \frac{2\Delta t\Lambda}{6}(v_{lca}(n) + \\ + 4v_{lca}(n-1) + v_{lca}(n-2)) \quad \forall n \in \{1, 2, \ldots, N\}
\end{gather*}

\subsection{Delta Load with Line-Ground Fault}

\begin{figure*}[!htbp]
\centering
\subfloat[Single-phase RL series load]{\includegraphics[width=1.35in]{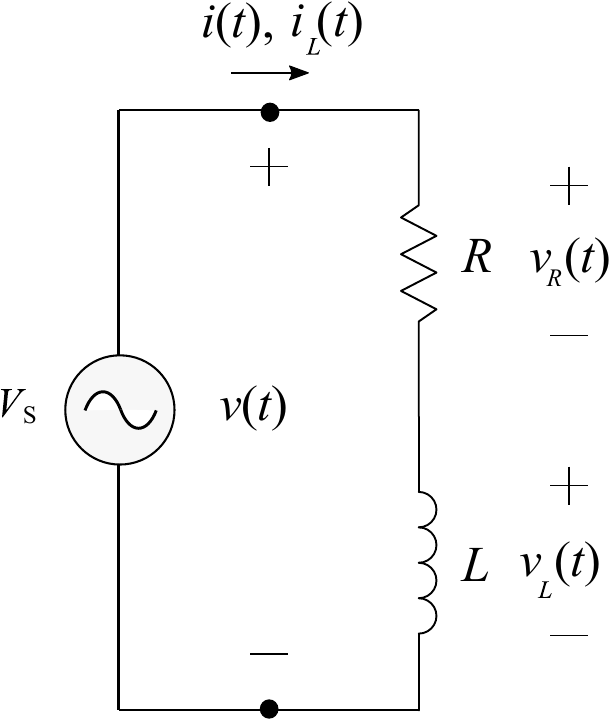}
\label{fig:rl-dynamic}}
\hfil
\subfloat[Grounded-wye-connected RL load]{\includegraphics[width=2.95in]{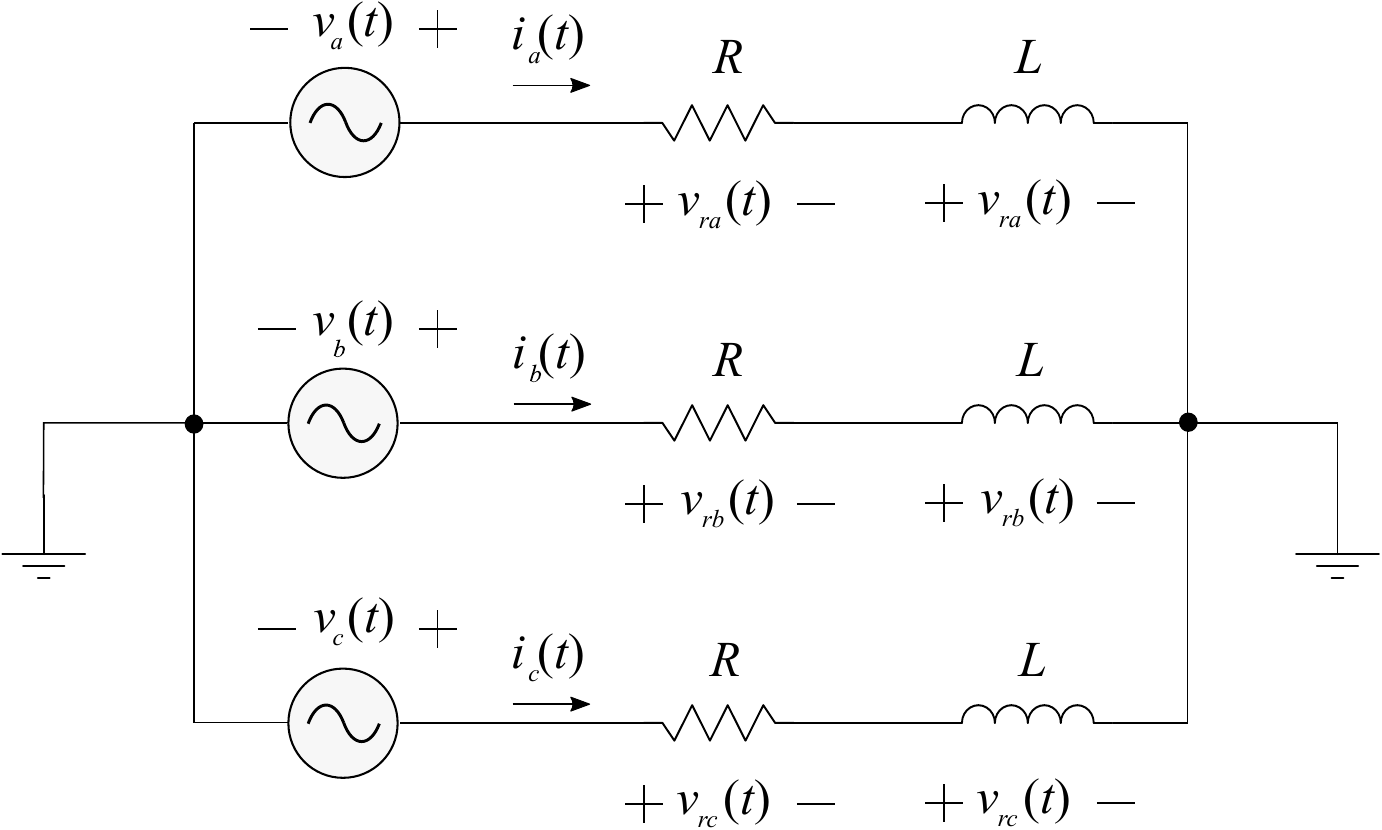}
\label{fig:gwye-no-fault-dynamic}}

\subfloat[Grounded-wye-connected RL load with a line-ground fault]{\includegraphics[width=2.95in]{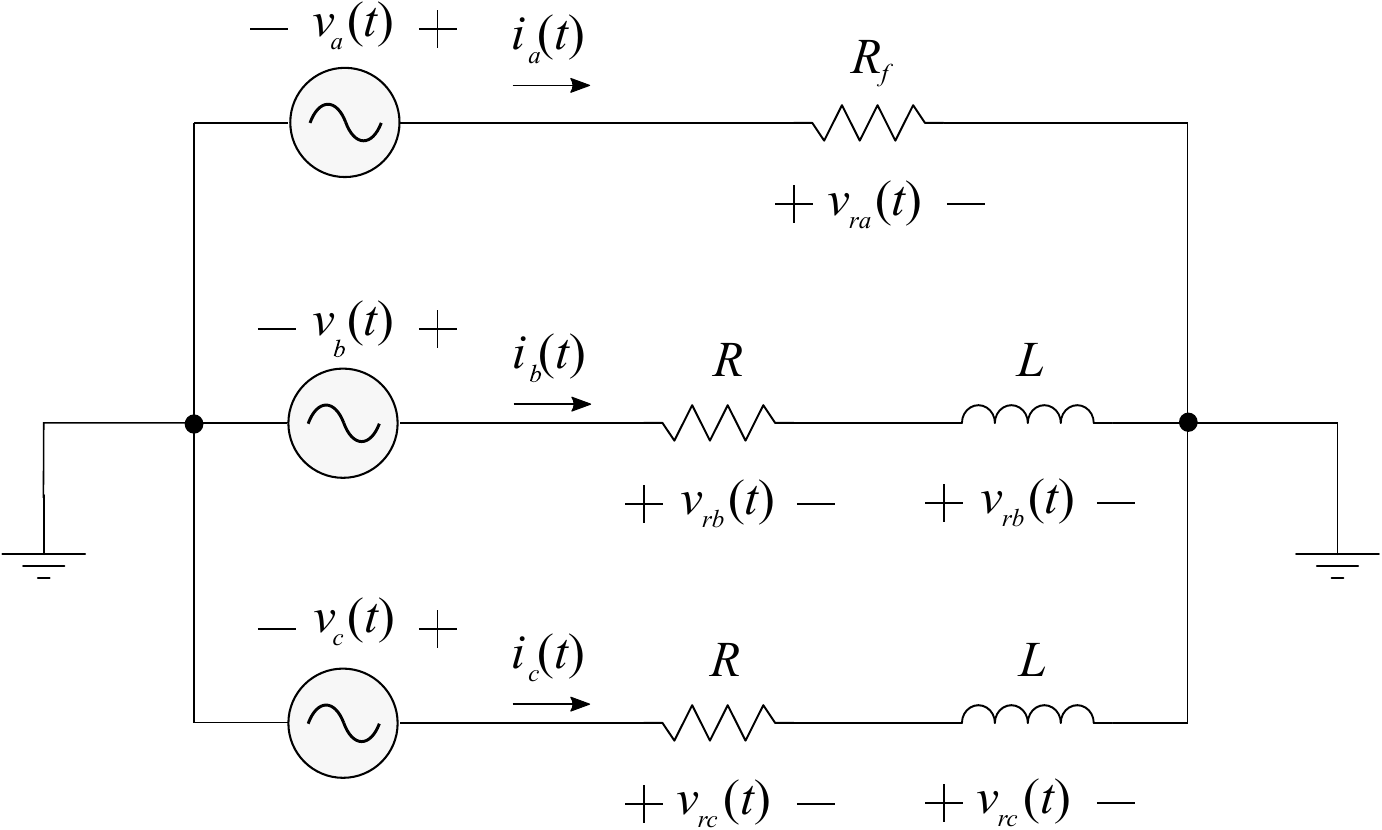}
\label{fig:gwye-lg-fault-r-only-dynamic}}
\hfil
\subfloat[Grounded-wye-connected RL load with a line-line fault]{\includegraphics[width=3.35in]{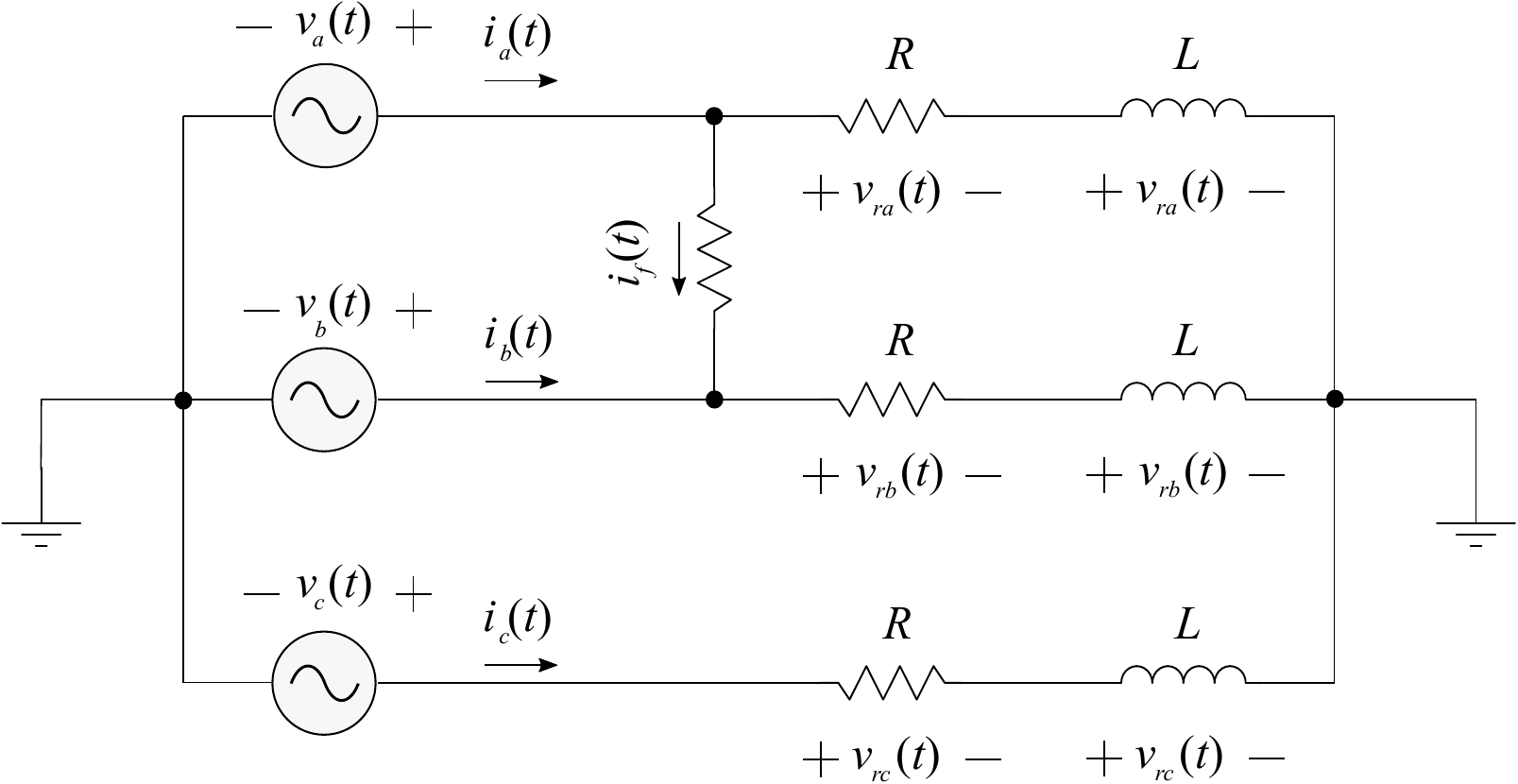}
\label{fig:gwye-ll-fault-r-only-dynamic}}

\subfloat[Delta-connected RL load]{\includegraphics[width=2.40in]{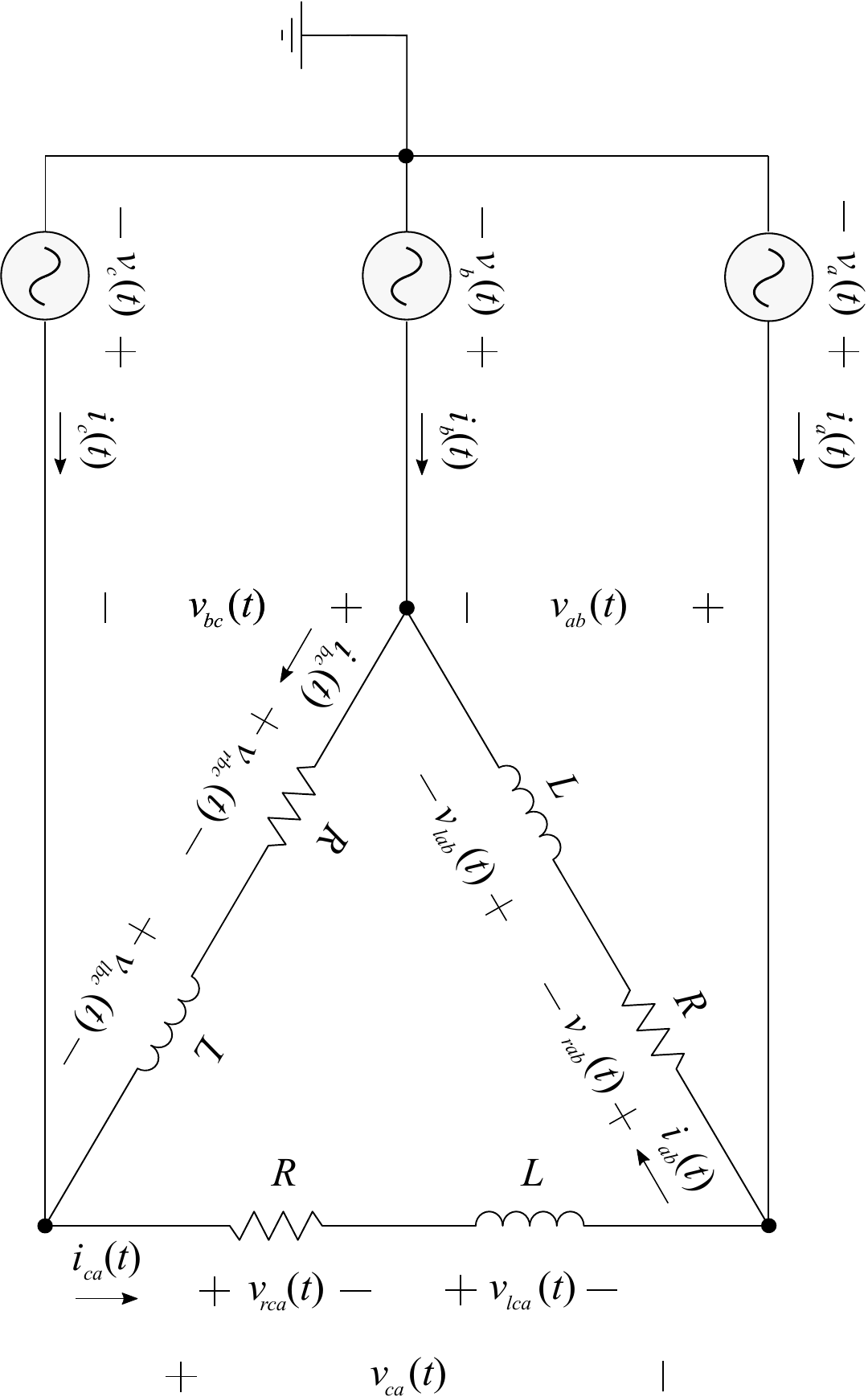}
\label{fig:delta-no-fault-dynamic}}
\hfil
\subfloat[Delta-connected RL load with a line-line fault]{\includegraphics[width=2.40in]{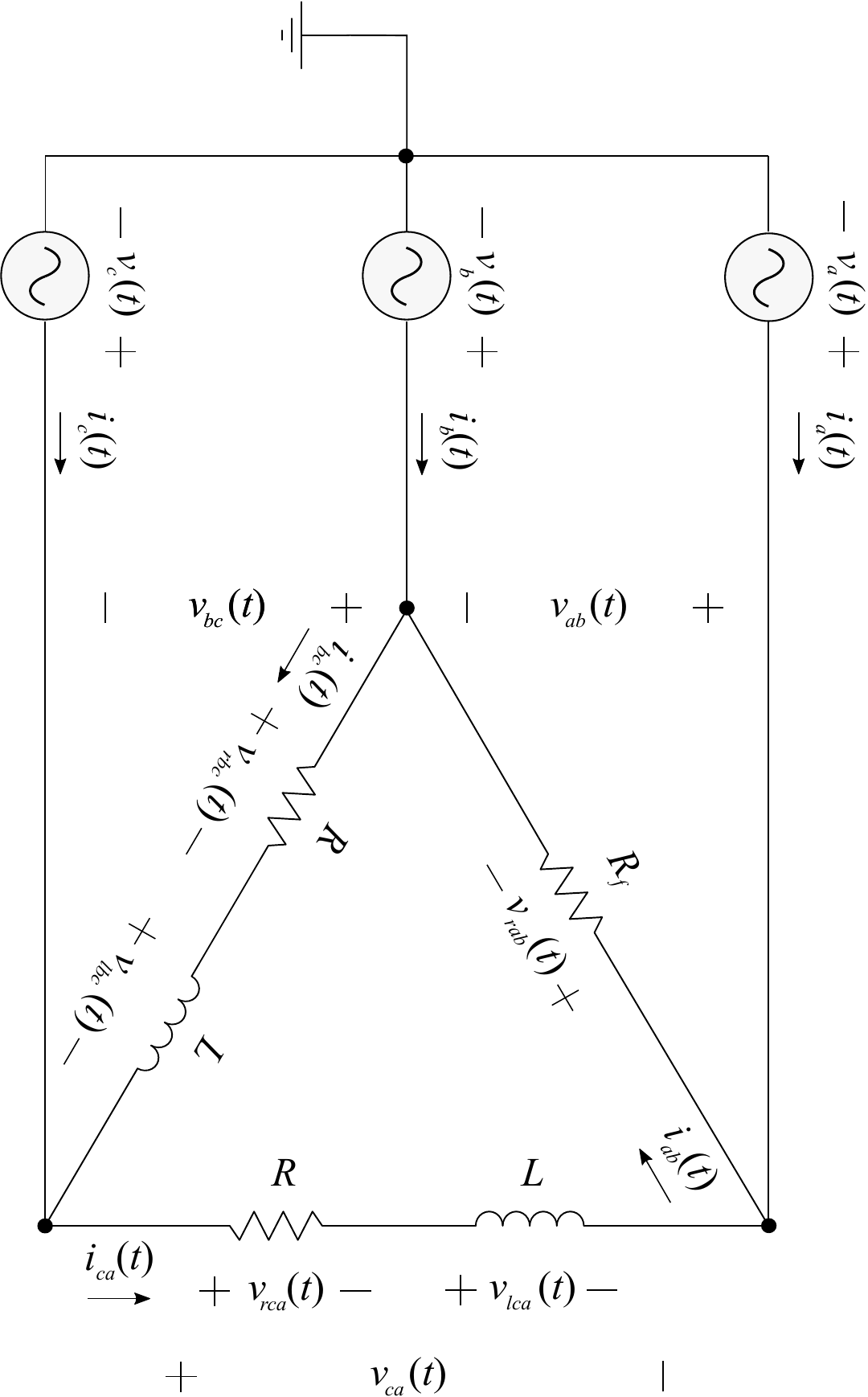}
\label{fig:delta-ll-fault-r-only-dynamic}}
\hfil
\subfloat[Delta-connected RL load with a line-ground fault]{\includegraphics[width=2.10in]{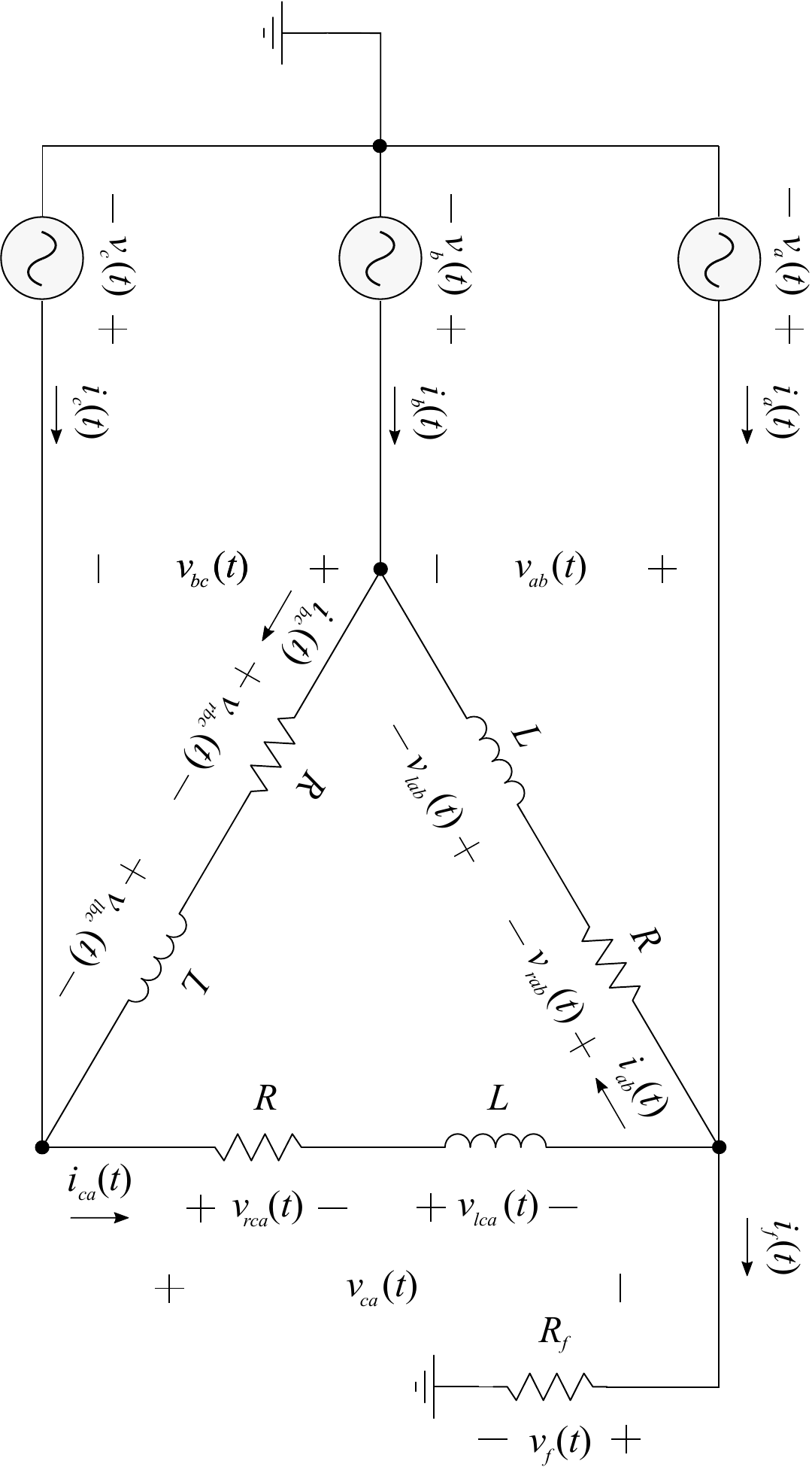}
\label{fig:delta-lg-fault-r-only-dynamic}}

\vspace{0.1in}
\caption{Dynamic implementation of state estimation-based protection, applied to single-phase, grounded-wye and delta-connected load configurations.}
\label{fig:dynamic-models}
\end{figure*}

The sampled output of the system (illustrated in Fig.~\ref{fig:delta-lg-fault-r-only-dynamic}):
\begin{equation*}
\mathbf{y} = \begin{bmatrix} \mathbf{v}_{ab} & \mathbf{v}_{bc} & \mathbf{v}_{ca} & \mathbf{v}_a & \mathbf{i}_a & \mathbf{i}_b & \mathbf{i}_c, &\mathbf{z}_{ab} & \mathbf{z}_{bc} & \mathbf{z}_{ca} \end{bmatrix}^T
\end{equation*}

The state of the system:
\begin{gather*}
x(t) = \begin{bmatrix} G & \Lambda & G_f & \mathbf{v}_{rab} & \mathbf{v}_{rbc} & \mathbf{v}_{rca} & ...\end{bmatrix}^T \\
\begin{bmatrix} ... & \mathbf{v}_{lab} & \mathbf{v}_{lbc} & \mathbf{v}_{lca} & \mathbf{v}_f \end{bmatrix}^T   
\end{gather*}
\noindent where $G_f = R^{-1}$ is the fault conductance, $\mathbf{v}_f$ is the voltage across the fault and the remaining states are the same as those in that of the delta no-fault state.
\vspace{0.1in}

The output function $\mathbf{h}(\mathbf{x})$ can be written as:
\begin{gather*}
h_n(\mathbf{x}) = v_{rab}(n) + v_{lab}(n) \quad \forall n \in \{1, 2, \ldots, N\} \\
h_{n+N}(\mathbf{x}) = v_{rbc}(n) + v_{lbc}(n) \quad \forall n \in \{1, 2, \ldots, N\} \\
h_{n+2N}(\mathbf{x}) = v_{rca}(n) + v_{lca}(n) \quad \forall n \in \{1, 2, \ldots, N\} \\
h_{n+3N}(\mathbf{x}) = v_f(n) \quad \forall n \in \{1, 2, \ldots, N\} \\
h_{n+4N}(\mathbf{x}) = G(v_{rab}(n) - v_{rca}(n)) + G_fv_f(n) \\ \quad \forall n \in \{1, 2, \ldots, N\} \\
h_{n+5N}(\mathbf{x}) = G(v_{rbc}(n) - v_{rab}(n))\quad \forall n \in \{1, 2, \ldots, N\} \\
h_{n+6N}(\mathbf{x}) = G(v_{rca}(n) - v_{rbc}(n)) \quad \forall n \in \{1, 2, \ldots, N\} \\    
h_{n+7N}(\mathbf{x}) = G(v_{rab}(n) - v_{rab}(n-2)) - \frac{2\Delta t\Lambda}{6}(v_{lab}(n) + \\ + 4v_{lab}(n-1) + v_{lab}(n-2)) \quad \forall n \in \{1, 2, \ldots, N\} \\
h_{n+8N}(\mathbf{x}) = G(v_{rbc}(n) - v_{rbc}(n-2)) - \frac{2\Delta t\Lambda}{6}(v_{lbc}(n) + \\ + 4v_{lbc}(n-1) + v_{lbc}(n-2)) \quad \forall n \in \{1, 2, \ldots, N\} \\
h_{n+9N}(\mathbf{x}) = G(v_{rca}(n) - v_{rca}(n-2)) - \frac{2\Delta t\Lambda}{6}(v_{lca}(n) + \\ + 4v_{lca}(n-1) + v_{lca}(n-2)) \quad \forall n \in \{1, 2, \ldots, N\}
\end{gather*}

\section{Experiments} \label{sec:experiments}
\indent

Three different case-study systems are considered: 1) a single-phase load, 2) a grounded-wye constant-impedance load, and 3) a delta-connected constant impedance load.
These load configurations are studied for both phasor-based state estimation and DSE. To verify the noise immunity of the methods, random noise with an amplitude of approximately 10\% of the signal peak is added to the measurements in both cases.
Experiments are performed in Julia~v1.5 \cite{julia} and 64-bit MATLAB~R2019b \cite{matlab}.

\subsection{Phasor Implementation} \label{sec:phasor-experiments}

For the single-phase phasor model, it is assumed that the source voltage is 240~[V] and the load impedance $R + jX$ is such that it draws a current of $10 - j5$~[A].
For the three-phase phasor models, both line-ground and line-line fault configurations are considered. These assume that the voltage source is 480~[V] rms line-line and the load impedance $R + jX$ is such that it draws $30 - i15$~[A] per phase.
The fault resistance $R_f$ is selected such that $R_f = R/10$.

Measured data is obtained by assuming a balanced input voltage and calculating the current by multiplying the input voltage phasor vector with the admittance matrix of the load-fault network.
This is also the case for the single-phase dynamic load, however, in that case the measured phasor voltage is converted to instantaneous voltage to obtain the input for DSE.

\subsection{Dynamic Implementation} \label{sec:dynamic-experiments}

The first case-study system is solved ad-hoc, assuming an ideal source with the parameters listed in Table~\ref{table:single-phase-dynamic-params}.

\vspace{0.1in}
\begin{table}[!htbp]
\begin{center}
\begin{tabular}{lcrl}
\hline
Variable & Symbol & Value & Units \\ 
\hline \hline
Total load real power & $P$ & 10 & kW \\
Total load reactive power & $Q$ & 5 & kVAR \\
Line-line RMS source voltage & $V_{ll}$ & 480 & V \\
Simulation time & T & 10 & ms \\
Sample rate & $T_s$ & 100 & $\mu$s \\
\hline
\end{tabular}
\end{center}
\caption{Parameters for Single-Phase Dynamic Load}
\label{table:single-phase-dynamic-params}
\end{table}

The latter two case-study systems are modeled in the MATLAB/Simulink\textsuperscript{\textregistered} SimScape multi-physics 
simulation environment, using the Specialized Power Systems library with the parameters listed in Table~\ref{table:three-phase-dynamic-common-params} and Table~\ref{table:three-phase-dynamic-varying-params}.

\vspace{0.1in}
\begin{table}[!htbp]
\begin{center}
\begin{tabular}{lcrl}
\hline
Variable & Symbol & Value & Units \\ 
\hline \hline
Total load real power & $P$ & 10 & kW \\
Total load reactive power & $Q$ & 5 & kVAR \\
Line-line RMS source voltage & $V_{ll}$ & 240 & V \\
Source resistance & $R_s$ & 19.2& $\Omega$ \\
Source inductance & $L_s$ & 25.465 & mH \\

Fault resistance & $R_f$ & 1 & m$\Omega$ \\
Ground resistance & $R_g$ & 10 & m$\Omega$ \\
Cable positive-sequence resistance & $R_c$ & 183.7 & m$\Omega$ \\
Cable positive-sequence reactance & $L_c$ & 26.6 & m$\Omega$ \\
Simulation time & T & 200 & ms \\
Fault start time & $T_f$ & 50 & ms \\
\hline
\end{tabular}
\end{center}
\caption{Common Parameters for Three-Phase Dynamic Models}
\label{table:three-phase-dynamic-common-params}
\end{table}

\begin{table}[!htbp]
\begin{center}
\begin{tabular}{lcrl}
\hline
Variable & Grounded-Wye & Delta \\ 
\hline \hline
Load resistance R ($\Omega$) & 18.432 & 55.296 \\
Load inductance L (mH) & 24.457 &  73.3 \\
\hline
\end{tabular}
\end{center}
\caption{Varying Parameters for Three-Phase Dynamic Models}
\label{table:three-phase-dynamic-varying-params}
\end{table}

In the systems, the load is connected to a 480~[V] rms line-line source through 1000~[ft] of 1/0 AWG quadruplex overhead service drop cable.
Three different cases are considered: 1) no-fault, 2) line-ground fault, and 3) line-line fault.

\subsection{Grounded-Wye Load}

\begin{figure}[!htbp]
\centering
\includegraphics[scale=0.30]{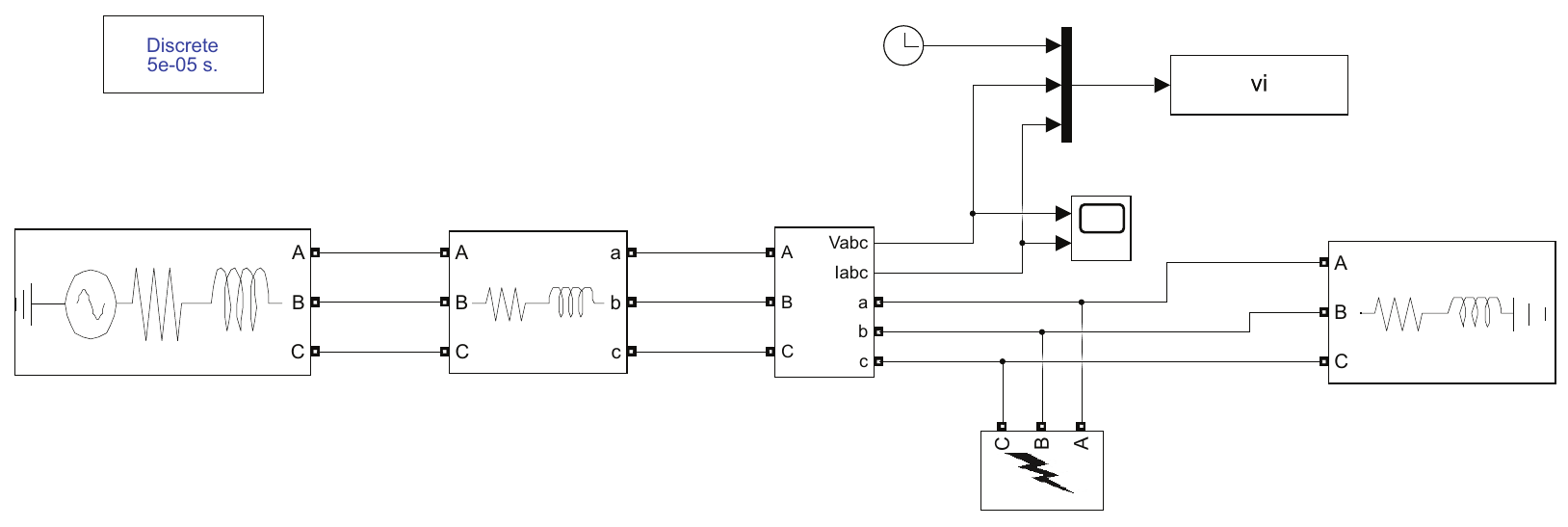}
\caption{MATLAB Simulink model for a grounded-wye load with faults.}
\label{fig:gwye-simulink}
\end{figure}

For the grounded-wye case, the system used is depicted in Fig.~\ref{fig:gwye-simulink}.
The load consists of three balanced series RL branches wired in a grounded-wye configuration.
This system has the parameters listed in Tables~\ref{table:three-phase-dynamic-common-params} and \ref{table:three-phase-dynamic-varying-params}. Note that the total fault resistance for the line-ground fault is $R_f + R_g = 110$~[m$\Omega$], while the total fault resistance for the line-line fault is $2R_f = 200$~[m$\Omega$].

\subsection{Delta Load}

\begin{figure}[!htbp]
\centering
\subfloat[Main model]{\includegraphics[width=3.25in]{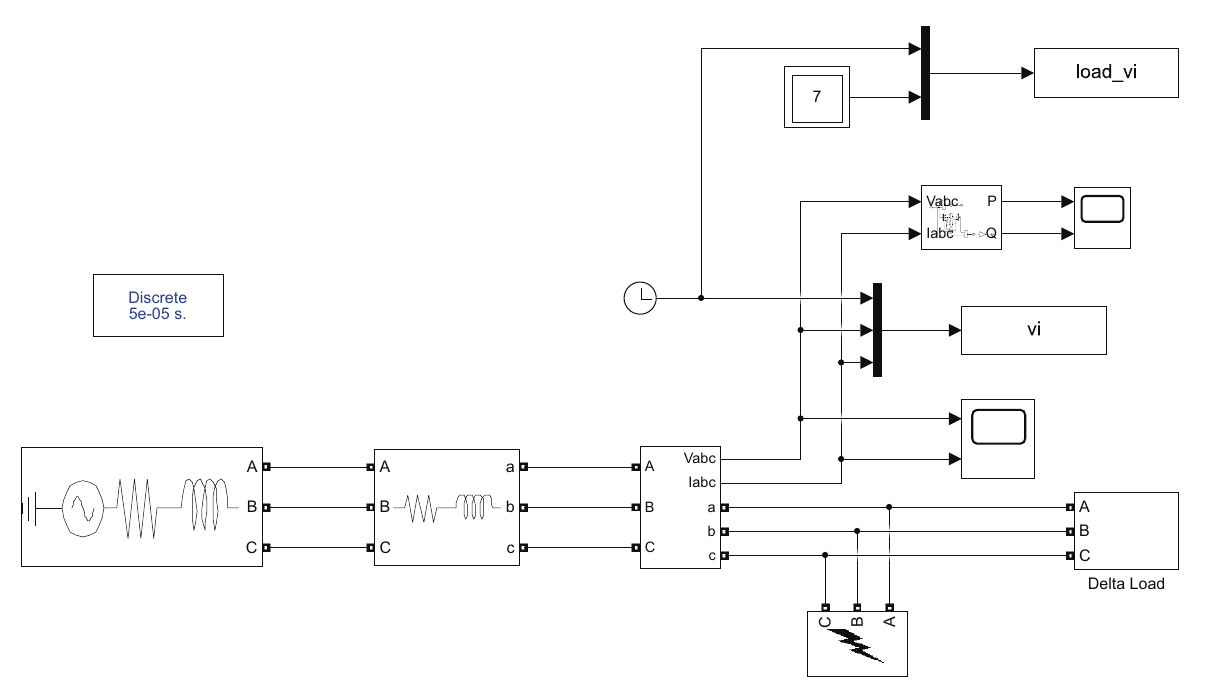}}
\hfil
\subfloat[Delta load]{\includegraphics[width=2.5in]{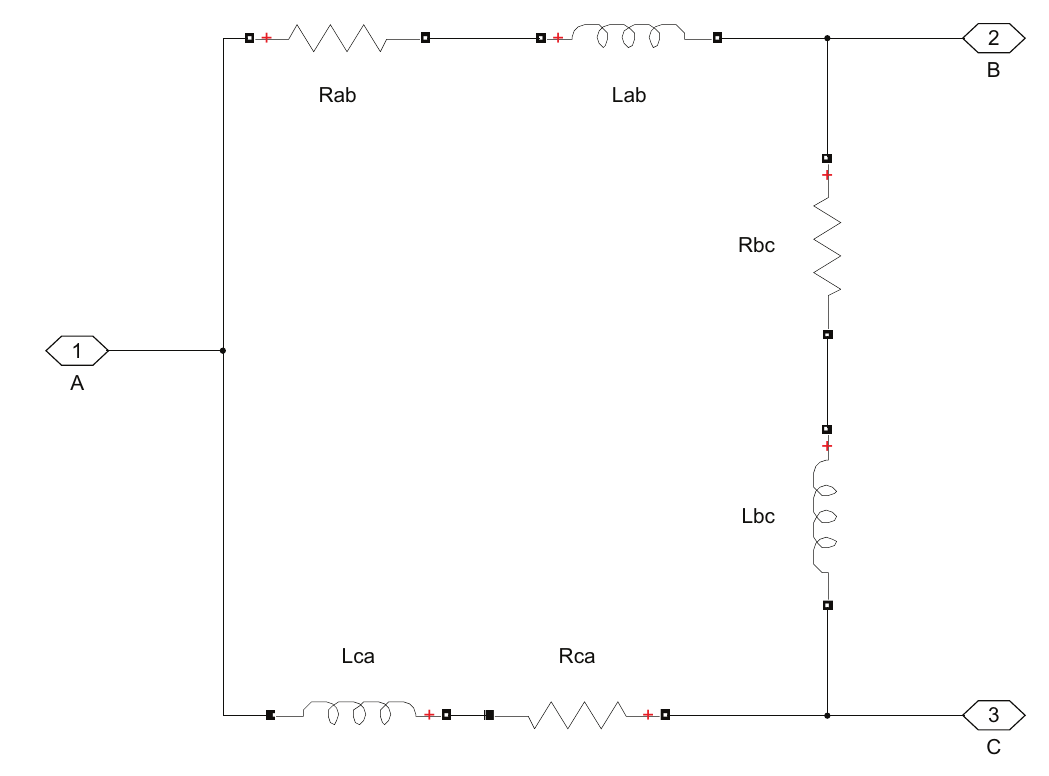}}
\vspace{0.1in}
\caption{MATLAB Simulink model for a delta load with faults.}
\label{fig:delta-simulink}
\end{figure}

For the delta-connected case, the system used is depicted in Fig.~\ref{fig:delta-simulink}.
The load consists of three balanced series RL branches wired in a delta configuration. 
This system has the parameters listed in Tables~\ref{table:three-phase-dynamic-common-params} and \ref{table:three-phase-dynamic-varying-params}. Note that the total fault resistance for the line-ground fault is $R_f + R_g = 110$~[m$\Omega$], while the total fault resistance for the line-line fault is $2R_f = 200$~[m$\Omega$].

\section{Results} \label{sec:results}
\indent

Results for the phasor models are presented in Table~\ref{table:phasor-results}, while results for the dynamic models are presented in Table~\ref{table:dynamic-results}.
The phasor models provide an excellent estimate of the system parameters.
DSE has difficulty estimating the fault resistance for the dynamic model of grounded-wye network with a line-line fault; a potential solution is to reduce the model order by neglecting the load impedance on the faulted phases.
Some moderate error is observed for the case of the delta-connected load with a line-ground fault; again, it may be possible to improve performance by neglecting load impedance on the faulted phases.

\begin{table*}[!htbp]
\begin{center}
\begin{tabular}{lrrrrrr}
\hline
Case & $R$ & $\hat{R}$ & $X$ & $\hat{X}$ & $R_f$ & $\hat{R}_f$ \\ 
\hline \hline
Single-Phase RL Load & 19.200 & 19.200 & 9.600 & 9.600 & -- & -- \\
Grounded-Wye Line-Ground Fault & 7.387 & 7.387 & 3.693 & 3.693 & 0.923 & 0.923 \\
Grounded-Wye Line-Line Fault & 7.387 & 5.184 & 3.693 & 4.787 & 0.923 &  0.935 \\
Delta Line-Line Fault & 22.160 & 22.160 & 11.080 & 11.080 & 2.770 & 2.770 \\
Delta Line-Ground Fault & 22.160 & 22.160 & 11.080 & 11.080 & 2.770 & 2.770 \\
\hline
\end{tabular}
\end{center}
\caption{Results for Phasor State Estimation}
\label{table:phasor-results}
\end{table*}

\begin{table*}[!htbp]
\begin{center}
\begin{tabular}{lrrrrrr}
\hline
Case & $R$ ($\Omega$) & $\hat{R}$ ($\Omega$) & $L$ (mH) & $\hat{L}$ (mH) & $R_f$ ($m\Omega$) & $\hat{R}_f$ ($m\Omega$) \\ 
\hline \hline
Single-Phase RL Load & 19.200 &  19.265 & 25.465 & 25.988 & -- & -- \\
Grounded-Wye No Fault & 18.432 & 18.404 & 24.446 & 24.485 & -- & -- \\
Grounded-Wye Line-Ground Fault & 18.432 & 18.432 & 24.446 & 24.446 & 11.000 & 10.997 \\
Grounded-Wye Line-Line Fault & 18.432 & 18.432 & 24.446 & 24.446 & 11.000 & 3.165 \\
Delta No Fault & 55.296 & 55.412 & 73.339 & 73.495 & -- & -- \\
Delta Line-Line Fault  & 55.296 & 55.895 & 73.339 & 73.666 & 2.000 & 2.001 \\
Delta Line-Ground Fault & 55.296 & 55.479 & 73.339 & 73.495 & 11.000 & 11.405 \\
\hline
\end{tabular}
\end{center}
\caption{Results for Dynamic State Estimation}
\label{table:dynamic-results}
\end{table*}

\section{Conclusions} \label{sec:conclusions}
\indent

The results of performed experiments prove that DSE is capable of correctly identifying model parameters of three different load configurations for both normal and faulted operations. These load configurations model a lumped equivalent of a radial electrical grid supplying multiple loads.

Several models showed sensitivity to inital conditions, particularly the delta-connected load, so it is important that consideration be given to providing the presented methods with good initial conditions.
One issue is making sure that there is a sufficient number of measurements to estimate model states; for example, it is not possible to infer impedances for an unbalanced delta-connected load given a single time snapshot.
To reduce the number of states, the models presented here assume that loads are balanced; this assumption can be an issue for systems with a high degree of load imbalance.

Existing work has demonstrated that DSE can operate with nonlinear elements \cite{albinali_state_2017}.
One option for future work is to expand the here presented methods with other load models. These could include nonlinear voltage-dependent models where power is a polynomial function of voltage (ZIP loads) or where power is a polynomial function of both voltage and frequency (e.g. the WSCC load model \cite{Pereira2002}).
Alternately, these could include dynamic load models (e.g. an induction motor model (MOTORW) or a composite load model (CMPLDW) \cite{nerc_dynamic_2016}).
Last, there is the possibility of protecting line sections that include loads with coordinated breakers at both ends; this could correspond to a distributed parameter line or a Pi/Tee lumped equivalent model \cite{rizy_evaluation_2002}.

%%%%% %%%%% %%%%%  - - - - - - - - - - - - -  %%%%% %%%%% %%%%%
%%%%%                      REFERENCES                     %%%%%

\newpage
\bibliographystyle{unsrt}
\bibliography{references}

\begin{thebibliography}{10}

\bibitem{meliopoulos_dynamic_2017}
A.~P.~S. {Meliopoulos}, G.~J. {Cokkinides}, P.~{Myrda}, Y.~{Liu}, R.~{Fan},
  L.~{Sun}, R.~{Huang}, and Z.~{Tan}.
\newblock {Dynamic State Estimation-Based Protection: Status and Promise}.
\newblock {\em {IEEE Transactions on Power Delivery}}, 32(1):320--330, Feb.
  2017.

\bibitem{liu_dynamic_2017}
Y.~{Liu}, A.~P.~S. {Meliopoulos}, R.~{Fan}, L.~{Sun}, and Z.~{Tan}.
\newblock {Dynamic State Estimation Based Protection on Series Compensated
  Transmission Lines}.
\newblock {\em {IEEE Transactions on Power Delivery}}, 32(5):2199--2209, 2017.

\bibitem{liu_dynamic_2016}
Y.~{Liu}, A.~P.~S. {Meliopoulos}, L.~{Sun}, and R.~{Fan}.
\newblock {Dynamic State Estimation Based Protection of Mutually Coupled
  Transmission Lines}.
\newblock {\em {CSEE Journal of Power and Energy Systems}}, 2(4):6--14, Dec.
  2016.

\bibitem{Liu2015b}
Y.~{Liu}, A.~P.~S. {Meliopoulos}, R.~{Fan}, and L.~{Sun}.
\newblock {Dynamic State Estimation Based Protection of Microgrid Circuits}.
\newblock In {\em {2015 IEEE Power Energy Society General Meeting}}, pages
  1--5, Jul. 2015.

\bibitem{Vasios2018}
O.~{Vasios}, S.~{Kampezidou}, and A.~P.~S. {Meliopoulos}.
\newblock {A Dynamic State Estimation Based Centralized Scheme for Microgrid
  Protection}.
\newblock In {\em {Proceedings of the 2018 North American Power Symposium}},
  pages 1--6, Sep. 2018.

\bibitem{Choi2017}
S.~{Choi} and A.~P.~S. {Meliopoulos}.
\newblock {Effective Real-Time Operation and Protection Scheme of Microgrids
  Using Distributed Dynamic State Estimation}.
\newblock {\em {IEEE Transactions on Power Delivery}}, 32(1):504--514, Feb.
  2017.

\bibitem{tumilty_approaches_2006}
R.~M. {Tumilty}, M.~{Brucoli}, G.~M. {Burt}, and T.~C. {Green}.
\newblock {Approaches to Network Protection for Inverter Dominated Electrical
  Distribution Systems}.
\newblock In {\em {Proceedings of the 3rd IET International Conference on Power
  Electronics, Machines and Drives, 2006}}, pages 622--626, Apr. 2006.

\bibitem{dewadasa_line_2008}
M.~{Dewadasa}, A.~{Ghosh}, and G.~{Ledwich}.
\newblock {Line Protection in Inverter Supplied Networks}.
\newblock In {\em {Proceedings of the 2008 Australasian Universities Power
  Engineering Conference}}, pages 1--6, Dec. 2008.

\bibitem{barnes21-admitrelay}
A.~K. {Barnes} and A.~{Mate}.
\newblock {Implementing Admittance Relaying for Microgrid Protection}.
\newblock In {\em {Proceedings of the 2021 IEEE/IAS 57th Industrial and
  Commercial Power Systems Technical Conference}}, pages 1--9, Apr. 2021.

\bibitem{dewadasa_distance_2008}
J.~M. {Dewadasa}, A.~{Ghosh}, and G.~{Ledwich}.
\newblock {Distance Protection Solution for a Converter Controlled Microgrid}.
\newblock In {\em {Proceedings of the 15th National Power Systems Conference}},
  2008.

\bibitem{Kothari1989}
D.~P. {Kothari} and I.~J. {Nagrath}.
\newblock {\em {Modern Power System Analysis}}.
\newblock {Tata McGraw-Hill Education}, 1989.

\bibitem{Monticelli2000}
A.~{Monticelli}.
\newblock {Electric Power System State Estimation}.
\newblock {\em {Proceedings of the IEEE}}, 88(2):262--282, Feb. 2000.

\bibitem{Chapra2009}
S.~{Chapra} and R.~{Canale}.
\newblock {\em {Numerical Methods for Engineers}}.
\newblock {McGraw-Hill Education}, 2009.

\bibitem{julia}
J.~{Bezanson}, S.~{Karpinski}, V.~B. {Shah}, and A.~{Edelman}.
\newblock {Julia:~A Fast Dynamic Language for Technical Computing}.
\newblock {\em arXiv:1209.5145}, 2012.

\bibitem{matlab}
{MathWorks}.
\newblock {MATLAB R2019b Documentation}.
\newblock {\em MATLAB -- https://www.mathworks.com}, 2020.

\bibitem{albinali_state_2017}
H.~F. {Albinali}.
\newblock {\em {State Estimation-Based Centralized Substation Protection
  Scheme}}.
\newblock PhD thesis, {Georgia Institute of Technology}, Aug. 2017.

\bibitem{Pereira2002}
L.~{Pereira}, D.~{Kosterev}, P.~{Mackin}, D.~{Davies}, J.~{Undrill}, and
  Z.~{Wenchun}.
\newblock {An Interim Dynamic Induction Motor Model for Stability Studies in
  the WSCC}.
\newblock {\em {IEEE Transactions on Power Systems}}, 17(4):1108--1115, Nov.
  2002.

\bibitem{nerc_dynamic_2016}
{North American Electric Reliability Corporation}.
\newblock {Technical Reference Document: Dynamic Load Modeling}.
\newblock Technical report, NERC, 2016.

\bibitem{rizy_evaluation_2002}
D.~T. {Rizy} and R.~H. {Staunton}.
\newblock {Evaluation of Distribution Analysis Software for DER Applications}.
\newblock Technical report, {Oak Ridge National Laboratory}, 2002.

\end{thebibliography}

\end{document}